# LYMAN $\alpha$ ABSORPTION AT LOW REDSHIFTS
# AND HOT GAS IN GALACTIC HALOES

H. J. Mo

Max-Planck-Institut für Astrophysik
Karl-Schwarzschild-Str. 1
85748 Garching bei München
Germany

**Abstract**

Motivated by recent observation of Lanzetta et al. that most luminous galaxies at low redshifts produce Ly$\alpha$ absorption at impact parameter $l \lesssim 160\,h^{-1}$kpc, we propose that these absorbers are clouds confined by the pressure of ambient hot gas in galactic haloes. We determine the properties of this hot gas and of the absorption systems on the basis of observational and theoretical constraints. The absorbing clouds need to be replenished on about one orbital time ($\sim 10^9$ yrs) in the galactic halo. The pressure and temperature of the gas at radius $r \sim 100\,$kpc are $P = (10 - 100)\,\text{cm}^{-3}\,$K, $T = 10^{(5.5-6.5)}\,$K. The model requires that about 10 per cent of the gas in low-redshift galactic haloes is in the hot phase. Such gas in galactic haloes emits x-ray with bolometric luminosity of the order $10^{37-40}\,\text{erg}\,\text{s}^{-1}$. The plausibility for such gas to exist in current models of galaxy formation is discussed.

## 1. Introduction

The study of hot gas in the haloes of normal galaxies is important for the understanding of galaxy formation (see e.g. White 1991 for a review). If dilute and cool (i.e. with $T \lesssim 10^6\,$K), the hot gas may be difficult to detect directly in x-ray emission. However, the QSO absorption systems associated with galaxies can be used to probe the properties of such tenuous gas. Recently, Lanzetta et al. (1994, hereafter LBTW) found that some luminous galaxies can produce Ly$\alpha$ absorptions at impact parameter as large as $r \sim 300\,h^{-1}$kpc, that most luminous galaxies produce Ly$\alpha$ absorptions at $r \lesssim 160\,h^{-1}$kpc. This conclusion confirms previous speculation that normal, luminous galaxies posses huge gaseous haloes (Bahcall & Spitzer 1969) or huge gaseous discs (Maloney 1992; Hoffman et al. 1993). Their results also offer the opportunity to study the properties of the component of tenuous gas associated with normal galaxies.

In this paper, we assume that the Ly$\alpha$ absorptions associated with galaxies are pro-duced by clouds pressure-confined by the ambient hot gas in galactic haloes. We determine the properties of this hot gas and of the clouds on the basis of observational and theoretical constraints. We discuss how plausible such a scenario is in current models of galaxy formation. Models in which Ly$\alpha$ clouds are pressure-confined by a general intergalactic medium (IGM) have been discussed by Sargent et al. (1980) and Ostriker & Ikeuchi (1983).

## 2. Constraints on the hot gas in galactic haloes

We assume that normal galaxies are embedded in dark haloes with spherical symmetry and isothermal-sphere density profile:

$$\rho(r) = \frac{v_{\text{cir}}^2}{4\pi G r^2}, \tag{1}$$

where $v_{\text{cir}}$ ($= v_{200} \times 200 \text{ km s}^{-1}$) is the circular velocity of the halo, $r$ ($= r_{100} \times 100 \text{ kpc}$) is the distance to the halo centre. We also assume that the proton number density $n$ and the temperature $T$ ($= T_6 \times 10^6$ K) of the gas in a halo have spherical symmetry. We model a cloud by a uniform sphere with proton number density $n_{\text{c}}$ and constant temperature $T_{\text{c}}$ ($= T_4 \times 10^4$ K). Under the assumption that the radius of a cloud ($R_{\text{c}}$) is much smaller than that of a halo, the pressure equilibrium of a cloud at a distance $r$ from the halo center gives

$$n_{\text{c}} T_{\text{c}} = n(r) T(r) \equiv P(r), \tag{2}$$

where $P(r)$ is the pressure of the halo gas at $r$. Assuming that the clouds are nearly completely photoionized by a constant UV background with flux $J$, using ionization equilibrium, we obtain the number density of neutral hydrogen in the cloud:

$$n_{\text{HI}} = 1.4 \times 10^2 n_{\text{c}}^2 T_{\text{c}}^{-3/4} J_{21}^{-1} \text{ cm}^{-3}, \tag{3}$$

where $J_{21} \equiv J/[10^{-21} \text{erg cm}^{-2} \text{ s}^{-1} \text{ sr}^{-1} \text{ Hz}^{-1}]$. Neglecting the geometry factor, the HI column density is then

$$N_{\text{HI}} = n_{\text{HI}} R_{\text{c}}. \tag{4}$$

We assume that the clouds are pressure confined. The condition that a cloud is gravitationally stable is $U/|W| > 2/3$, where $U \approx 4\pi n_c R_c^3 k T_c$ is the internal energy, $W = -(3G/5)(4\pi/3)^2 m_p^2 n_c^2 R_c^5$ (with $m_p$ being the proton mass) is the self gravitational energy, of the cloud (Spitzer, 1978, p241). We have neglected the contribution of the dark matter to the gravitational energy. The dark matter potential tends to disrupt the clouds through tidal forces, as discussed later. The above condition gives

$$P(r) > 0.3 \times N_{14}^{2/3} T_4^{7/6} J_{21}^{2/3} \,\mathrm{K\,cm^{-3}}, \tag{5}$$

where $N_{14} \equiv N_{\mathrm{HI}}/(10^{14}\mathrm{cm}^{-2})$.

If clouds are destroyed on a time scale $t_b = t_{10} \times 10^{10}$ yrs, they must be replenished on a similar time scale. The value of $t_b$ is not known. If clouds are destroyed during the life of a galaxy, then $t_b < t_U$, where $t_U$ is the age of the universe. We also expect that $t_b$ is larger than the internal dynamical time of the clouds: $\tau_d \sim (m_p/3kT_c)^{1/2} R_c \approx 4\times 10^8 (R_c/10\,\mathrm{kpc}) T_4^{-1/2}$ yrs. In the following, we discuss several processes that can destroy clouds. The corresponding time scales give an upper limit on $t_b$.

The collision among clouds may destroy them. If they are moving around at about the sound speed of the hot gas, the time scale for the collision is about $t_{co} = 10^9 r_{100} T_6^{-1/2} f_c^{-1}(r)$ yrs, where $f_c(r)$ is the cloud covering factor within a radius $r$. The observation of LBTW shows that $f_c \sim 1$ for $r \lesssim 100$ kpc. If the hot gas is at the virial temperature of the galactic halo, $T_{\mathrm{vir}} \approx 1.4 \times 10^6 v_{200}^2$ K, the collision can destroy clouds on a time scale $\sim 10^9$ yrs. The collision will be less frequent, if the buoyancy force can prevent the clouds from moving at the sound speed of the hot gas. However, the data of LBTW shows that the velocities of the absorption systems with respect to the host galaxies are typically 100 - 200 km s$^{-1}$, which is about the sound speed in a hot gas at $T = 10^6$ K.

Clouds moving in gaseous haloes also suffer from friction due to the ram pressure of the hot gas. If the density of the hot gas is too high, this friction will cause the clouds to spiral to the centre of the galaxy, and then to fall into the disk. Assuming that the mass of the cloud $M_c$ is independent of time, and that the gas has the same density profile as the dark matter (eq.1), we obtain that a cloud reaches the center after a time $t_{\rm fr} = (2\pi R_c/v_{\rm cir})(T/T_c)$. It is clear that this process is less effective in disrupting clouds than the cloud-cloud collision, unless $R_c$ is small. The lower limit on $t_{\rm fr}$ is the free-fall time in the halo, which is about $6 \times 10^8 r_{100} v_{200}^{-1}$ yrs.

Clouds with radii larger than the tidal radius, or with densities $n_c \lesssim 0.01 \, r_{100}^2 v_{200}^{-2} {\rm cm}^{-3}$, can be disrupted by the tidal force of the dark halo. As discussed in Kurth (1957, p89), the tidal disruption will occur in roughly one dynamical time: $t_{\rm dyn} \approx 10^9 r_{100} v_{200}^{-1}$ yrs.

Since in our model the clouds are moving in a more diffuse medium, the clouds suffer Kelvin-Helmholtz (KH) instability. According to the result of Murray et al. (1993), pressure confined clouds moving at the sound speed of the medium are destroyed on a time scale $t_{\rm KH}$ which is about two times the internal dynamical time scale ($\tau_d$) of the cloud. For a cloud with radius $R_c$, $t_{\rm KH} \sim 10^9 (R_c/10{\rm kpc}) T_4^{-1/2}$ yrs. If $R_c \ll 10$ kpc, the KH instability can disrupt a cloud on a time scale much shorter than those given by the cloud-cloud collision and the tidal disruption. However, if clouds are initially at rest with respect to the medium, the time for them to be accelerated to the sound speed of the medium is about the dynamical time scale. The KH instability cannot destroy clouds on a time scale much shorter than this.

Based on the above considerations, we expect that the upper limit of $t_b$ is about one dynamical time of the cloud in the halo. That is $t_b \lesssim 10^9$ yrs. This conclusion does not depend strongly on the properties of the hot gas. This upper limit corresponds to a lower limit of the replenishment rate of the clouds, as discussed in section 3.

Clouds in a hot medium may also be disrupted by evaporation. In our case, the saturation parameter, which is essentially the ratio of the electron mean-free path to the cloud radius $R_c$ (Cowie & McKee 1977; Balbus & McKee 1982), $\sigma_0 = 1.2 \times 10^4 T^2 n^{-1} R_c^{-1} \lesssim 1$. The evaporation time scale in this case is $t_{\rm evp} \approx 1.1 \times 10^{-9} (n_c R_c)^2 T_c n^{-1} T^{-7/2}$ s. The condition that $t_{\text{evp}} > t_{\text{b}}$ gives

$$P(r)t_{10}^{1/3} < 1.2 N_{14}^{2/3} T_4^{3/2} J_{21}^{2/3} T_6^{-5/6} \text{cm}^{-3}\text{K}. \qquad (6)$$

Relation (6) can be considered as a constraint on the density and temperature of the hot gas when $t_{\text{b}}$ is given, or vice versa. Note that the dependence on $t_{\text{b}}$ is rather week.

Hot gas (i.e. gas with $T \gtrsim 10^4$) can cool if its density is high. The cooling time scale is $t_{\text{cool}} \approx 3kT/[n\Lambda(T)]$, where $\Lambda(T)$ is the cooling function (see Binney & Tremaine 1987, p580). The condition that $t_{\text{cool}}$ is larger than $t_{\text{U}}$ leads to

$$P(r) < 2.1 \times 10^2 T_6^2(r) \Lambda_{23}^{-1} h \text{cm}^{-3}\text{K}, \qquad (7)$$

where $\Lambda_{23} = \Lambda(T)/(10^{-23}\,\text{erg cm}^3\,\text{s}^{-1})$, $h$ is the Hubble constant in unit of $100\,\text{km s}^{-1}\,\text{Mpc}^{-1}$.

In Figure 1 we show the allowed region in $n - T$ plane given by the constraints (5, 6, 7). We take $v_{200} = 1$ for normal galaxies, $J_{21} = 0.05$ (see e.g. Miralda-Escude & Ostriker 1990; Songaila et al. 1989), $T_4 = 3$ for photoionization heating, and $N_{14} = 160$ for $r_{100} = 1$ (see LBTW). We also take $t_{\text{b}} = 10^9$ yrs. The two cooling curves (dashed lines) correspond to equation (7) with $h = 1$, assuming unenriched gas and gas with solar metallicity, respectively. If the column density is related to $r$ by $N_{14} \propto r_{100}^{-\alpha}$, then the constraints for another radius $r$ can be obtained according to their dependence on $r$ and $N_{14}$. The figure shows that, for unenriched gas, the temperature at $r \sim 100\,\text{kpc}$ is at the level of $10^{5.5-6.5}$ K. The temperature may be slightly higher, if the gas has solar metallicity. The pressure at this radius is about $(10\text{-}100)\text{Kcm}^{-3}$. The pressure at other radii depends on the value of $\alpha$ and on how the temperature $T$ of the hot gas changes with $r$. For a constant $T$, $\alpha = 5$ (as is given by LBTW) gives $P(r) \propto r^{-3}$, while $\alpha = 3$ leads to $P(r) \propto r^{-2}$. The statistics from the data is, however, not yet conclusive.

If the gas in a halo were constantly heated by accretions of material in IGM, and/or by supernova explosions, the constraint based on the cooling argument would be weaker than that given by equation (7), and the density of hot gas could be higher. However, hot gas in a galactic halo will emit x-ray. If the hot gas in a galactic halo has a constant temperature and a distribution which is similar to that of the dark matter (eq.1), then the total bolometric luminosity of the gas is of the order

$$L_{\rm x} \sim 10^{48} n^2(r) r_{100}^3 (r/r_{\rm disk}) \Lambda_{23} \text{ erg}, \tag{8}$$

where the core radius of the gas density is assumed to be the disk size $r_{\rm disk}$, which is taken to be 20 kpc. In figure 1, the lower and upper dotted lines show the boundaries on which $L_{\rm x} = 10^{40} \,{\rm erg\,s^{-1}}$ for unenriched gas and gas with solar metallicity, respectively. If the observed value of $L_{\rm x}$ for galaxies with circular velocity $v_{\rm cir} = 200 \,{\rm km\,s^{-1}}$ is indeed smaller than about $10^{40} \,{\rm erg\,s^{-1}}$ (e.g. McCammon & Sanders 1984), then the constraint on the density of the hot gas is slightly more stringent than that based on the cooling argument. This means that, if the hot gas has the low density as given by the x-ray observation, then it should not have been heated constantly, unless the x-ray emission can be reduced due to other reasons.

If the cloud replenishement occurs on a time scale much smaller than $10^9$ yrs, the constraint given by the evaporation is weaker than that shown in the figure. The pressure of the hot gas can be much higher. However, if the gas is not heated to a temperature much higher than the virial temperature of the halo, $T_{\rm vir} = 1.4 \times 10^6 v_{200}^2$ K, and if the constraint given by the x-ray observation is correct, then the pressure range given above is still approximately correct.

If most of the clouds producing the Ly$\alpha$ absorptions were in free expansion, the pressure could be smaller. The above pressure corresponds to a magnetic field of the order $\sim 0.3 \mu$G. It is not known whether or not such a magnetic field exists in galactic haloes to such a large radius as considered here. The magnetic field in the interstellar medium is of the order $1\mu$G. The pressure in the solar neighbourhood is about $3 \times 10^3 \text{ cm}^3\text{K}$. If clouds were made from material in the interstellar medium, adiabatically expanding to the pressure discussed above, then the magnetic field in clouds is of the order $0.1\mu$G. In this case, the pressure in the hot gas may be higher than that given above. However, the magnetic field in clouds should be much lower, if they are made from material in the halo or in the IGM.

## 3. The properties of absorption systems

The radius of a cloud can be written as (eq.4):

$$R_{\rm c} = \frac{N_{\rm HI}}{n_{\rm HI}} \approx 7.4 \times 10^{22} N_{14}(r) J_{21} T_4^{11/4} P^{-2}(r) \,{\rm cm}. \tag{9}$$

Taking $J_{21} = 0.05$, $T_4 = 3$, we find $R_{\rm c} \sim (1-10)\,{\rm kpc}$ for $r \lesssim 100\,{\rm kpc}$. To achieve a unit covering factor within this radius (LBTW), the number of clouds is of the order of $10^{2-4}$. The mass of each cloud $M_{\rm c} \sim 10^{5-7}\,M_\odot$, and the total gas mass in all clouds is about $10^9\,M_\odot$. If the cloud replenishment occurs on a time scale of about $10^9$ yrs, this total mass implies a replenishment rate of about $1\,M_\odot$ per year. The volume filling factor of the clouds is about 0.1. Assuming that the volume filling factor of the hot gas is 0.9, and using pressure equilibrium, we get that the total mass in the hot gas is about 0.3 times that in the clouds.

In the pressure-confinement model of Ostriker & Ikeuchi (1983), the expansion of the intergalactic medium may cause a rapid decrease in the comoving number density of Ly$\alpha$ absorbers with decreasing redshift, as is required by the observations at high redshifts (e.g. Lu et al. 1991). In our model, the inter-cloud medium is assumed to be stable, because it is confined by the static potential wells of galactic haloes, the evolution of the comoving number density of absorbers might be less stronger. *Hubble Space Telescope* data indicate that the evolution of the rate of incidence of low-redshift Ly$\alpha$ absorptions is indeed weak (Bahcall et al. 1991; Morris et al. 1991). However, this can also be a result of fortuitous cancellation of the effects of various things that are evolving.

The HI column density defined in equation (4) can be, using equations (2, 4), written

as

$$N_{\rm HI} \approx 4.1 \times 10^{13} \left(\frac{R_{\rm c}}{10\,{\rm kpc}}\right) P^2(r) T_4^{-11/4} J_{21}^{-1}\,{\rm cm}^{-2}, \qquad (10)$$

(Mo & Morris 1994). If the hot gas in a halo has a constant temperature, and if its distribution is similar to that given by eq.(1), then $P(r) \propto r^{-2}$. In the case where the cloud mass $M_{\rm c}$ scales with $r$ as $M_{\rm c} \propto r^{-\mu}$, we have $N_{\rm HI} \propto r^{-\alpha}$ with $\alpha = (10 + \mu)/3$. If $\mu = 0$, i.e. $M_{\rm c}$ is independent of $r$, then $\alpha = 10/3$. This value of $\alpha$ is slightly smaller than that given by the initial results of LBTW. Better data are needed to determine $\alpha$ more accurately. If the covering factor of absorbers in a halo is independent of $r$, then the relation $N_{\rm HI} \propto r^{-\alpha}$ leads to a column density distribution $f(N_{\rm HI}){\rm d}N_{\rm HI} \propto N_{\rm HI}^{-\beta} {\rm d}N_{\rm HI}$ with $\beta = (2+\alpha)/\alpha$. For $\alpha = 10/3$, we have $\beta = 1.6$. This value of $\beta$ is in good agreement with that observed for the Ly$\alpha$ absorption systems at $z \gtrsim 2$ (see Weymann 1993 for a review). The value of $\beta$ for low-redshift systems is still uncertain.

## 4. Discussion

In a universe with critical density, the virial radius and temperature of a halo at present time are $r_{\rm vir} \approx 200 v_{200}\, h^{-1}{\rm kpc}$, $T \approx 1 \times 10^6 v_{200}^2\,{\rm K}$, respectively. If all hot gas were to stay at the virial temperature, the pressure of the gas would be $P \approx 400(\Omega_{\rm b}/0.05) v_{200}^4 r_{100}^{-2}\,{\rm K\,cm}^{-3}$ (where $\Omega_{\rm b}$ is the cosmic mass density in baryon). This pressure is about 10 times larger than that discussed in section 2. So our model requires that only about 10 per cent of the baryons in galactic haloes stay in the hot phase. Recent simulations of Navarro & White (1994) show that, although galactic haloes are filled with hot gas at the virial temperature, only a small fraction (about 10 per cent) of gas ends up in the hot phase. If this were the case, then the existence of the hot gas required in our model would indeed be plausible in current models of galaxy formation. It is clear that more detailed simulations in the future will give more stringent constraints.

What is the origin of the clouds in galactic haloes? In the hierarchical scenario of structure formation, there are several possibilities. The merger of minihaloes to large systems may be quite frequent in current models of structure formation (e.g. Mo, Miralda-Escude & Rees 1993). The gas originally in a minihalo may be stripped out quickly by the ram-pressure of hot gas as the minihalo merges into a galactic halo. The pressure in consideration implies a gas density in clouds of the order $n_c \sim 10^{-3}\,\mathrm{cm}^{-3}$. For minihaloes with $v_{\mathrm{cir}} = 30\,\mathrm{km\,s}^{-1}$ (Rees 1986) at present time, the radius within which the baryon number density is equal to this value of $n_c$ is about 5 kpc, which is comparable to the value of $R_c$ discussed in section 3. So the inner parts of minihaloes may become pressure confined in galactic haloes. The interaction of two or more galaxies may produce tidal debrits, like the Magellannic Stream (MS), which may give rise to Ly$\alpha$ absorptions (Morris & van den Berg 1994). It is not clear whether or not these debrits are pressure confined or in free expansion. Morris & van den Berg argued that some kind of confinement may be needed for the debrits to exist long enough to produce the observed rate of incidence of the low-redshift Ly$\alpha$ absorptions. Recently, Moore & Davis (1994) proposed a model for the MS, which requires the existence of an extended dilute halo of diffuse hot gas surrounding the Milky Way. This gas will certainly interact with clouds that produce the Ly$\alpha$ absorption. It is interesting to note that the density of hot gas they obtained for large radii ($r \gtrsim 50\,\mathrm{kpc}$) is compatible with that required in our model. Furthermore, star formation in galaxies and thermal instabilities of the hot medium may also produce some inhomogeneities in galactic haloes. It would be interesting to see if these mechanisms will provide the right time scale for cloud productions.

## Acknowledgements

The author thanks Andreas Burkert, Martin Haehnelt, especially Jordi Miralda-Escude and Simon White for discussions and suggestions. He also thanks the referee, Cedric Lacey, for very helpful reports.

# References

Bahcall J.N., Jannuzi B.T., Schneider D.P., Hartig G.F., Bohlin R., Junkkarinen V., 1991, ApJ, 377, L5

Bahcall J.N., Spitzer L., 1969, ApJ, 156, L63

Balbus S.A., McKee C.F., 1982, ApJ, 252, 529

Binney J., Tremaine S.D., 1987, Galactic Dynamics, Princeton Univ. Press: Princeton

Cowie L.L., McKee C.F., 1977, ApJ, 211, 135

Hoffman, G. L., Lu, N. Y., Salpeter, E. E., Farhat, B., Lamphier, C. and Roos, T., 1993, AJ, 106, 39

Kurth R., 1957, Introduction to the Mechanics of Stellar Systems, Pergamon Press: New York

Lanzetta K., Bowen D.V., Tytler D., Webb J.K., 1994, ApJ, in press (LBTW)

Lu L., Wolfe A.M., Turnshek D.A., 1991, ApJ, 367, 19

Maloney P., 1992, ApJ, 398, L89

McCammon D., Sanders W.T., 1984, ApJ, 287, 167

Miralda-Escude J., Ostriker J.P., 1990, ApJ, 350, 1

Mo H.J., Miralda-Escude J., Rees M.J., 1993, MNRAS, 264, 705

Mo H.J., Morris S.L., 1994, MNRAS, in press

Moore B., Davis M., 1994, MNRAS, submitted

Morris S.L., Weymann R.J., Savage B.D., Gilliland R.L., 1991, ApJ, 377, L21

Morris S.L., van den Berg S., 1994, ApJ, in press

Murray S.D., White S.D.M., Blondin J.M., Lin D.N.C., 1993, ApJ, 407, 588

Navarro J., White S.D.M., 1994, MNRAS, 267, 401

Ostriker J.P., Ikeuchi S., 1983, ApJ, 263, L63

Rees M.J., 1986, MNRAS, 218, 25p

Sargent W.L.W., Young P.J., Bokenberg A., Tytler D., 1980, ApJS, 42, 41

Songaila A., Bryant W., Cowie L.L., 1989, 345, L71

Spitzer L., 1978, Physical Processes in the Interstellar Medium, Wiley-Interscience: New York

Weymann R.J., 1993, in Shull J.M., Thronson H.A., eds, The Environment and Evolution of Galaxies, Kluwer: Dordrecht, p213

White S.D.M., 1991, in Combes F., Casoli F., eds, Dynamics of Galaxies and Their Molecular Cloud Distributions, Kluwer: Dordrecht, p383

# Figure captions

**Figure 1** The properties of the hot gas in a galactic halo with circular velocity $v_{\rm cir} = 200\,{\rm km\,s^{-1}}$ determined by observational and theoretical constraints. The allowed regions in the $(n, T)$ plane are shown for hot gas at a distance $100\,{\rm kpc}$ from the halo center. The clouds are assumed to be created on a time scale $t_{\rm b} = 10^9\,{\rm yrs}$. The two solid lines show the constraints given by equations (5, 6), respectively. The dashed lines sketches the cooling curves (eq.7) for unenriched gas and gas with solar metallicity, respectively. The lower and upper dotted lines show the boundaries on which $L_{\rm x} = 10^{40}\,{\rm erg\,s^{-1}}$ (see eq.8) for unenriched gas and gas with solar metallicity, respectively. The horizontal line shows the virial temperature of the halo.

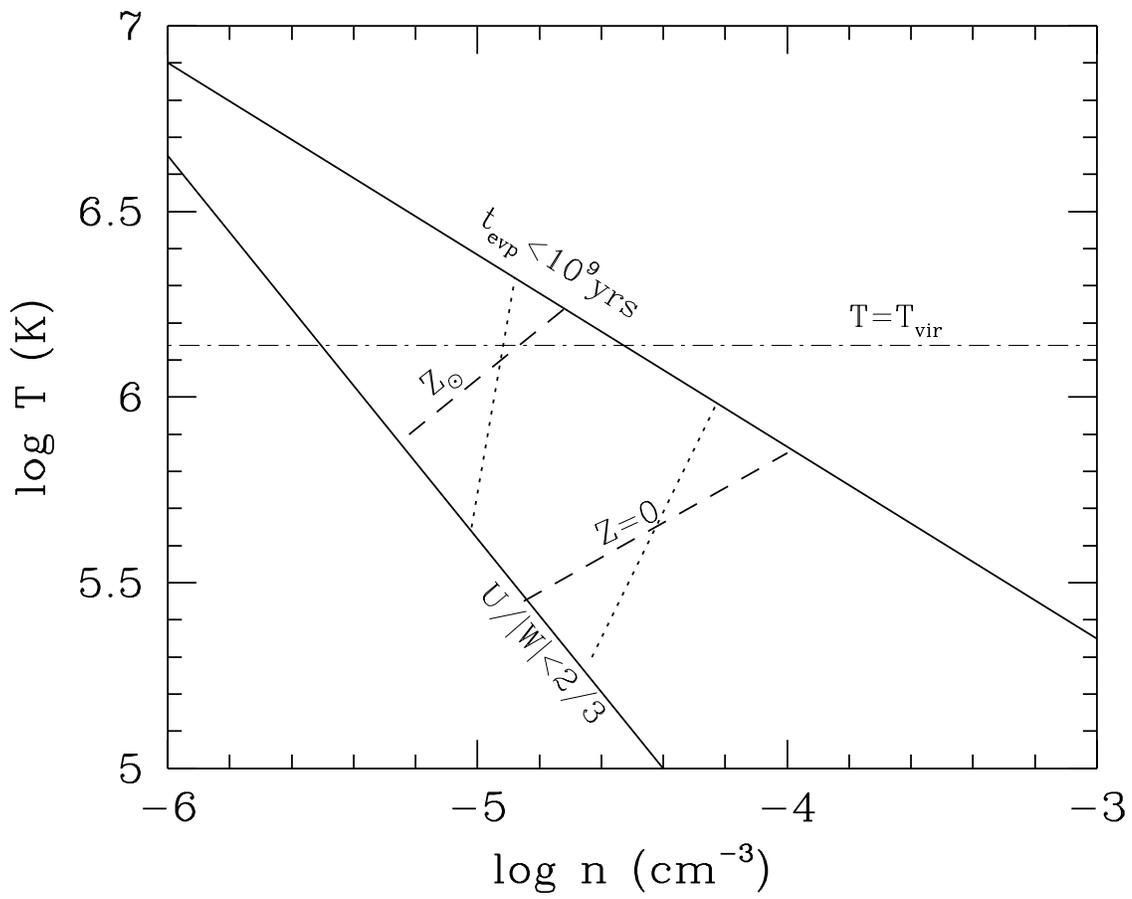